\documentclass[prc,aps,nofootinbib,showkeys,showpacs,twocolumn]{revtex4} 

\usepackage{amsmath}
\usepackage{amssymb}
\usepackage{amsfonts}
\usepackage{color}

\usepackage{epsfig}
\usepackage{graphicx}
\begin{document}
\title{Quantal corrections to mean-field dynamics including pairing
}
 
\author{Denis Lacroix} \email{lacroix@ganil.fr}
\affiliation{GANIL, CEA and IN2P3, Bo\^ite Postale 5027, 14076 Caen Cedex, France}
\author{Danilo Gambacurta} 
\affiliation{GANIL, CEA and IN2P3, Bo\^ite Postale 5027, 14076 Caen Cedex, France}

\author{Sakir Ayik}

\affiliation{Physics Department, Tennessee Technological University, Cookeville, TN 38505, USA}

\begin{abstract}  
Extending the stochastic mean-field model by including pairing, an approach is proposed for describing evolutions of complex many-body systems in terms of an ensemble of Time-Dependent Hartree-Fock Bogoliubov trajectories which is determined by incorporating fluctuations in the initial state. Non-linear evolution of the initial fluctuations provides an approximate description of quantal correlations and fluctuations of collective observables. Since the initial-state fluctuations break the particle-number symmetry, the dynamical description in which pairing correlations play a crucial role is greatly improved as compare to the mean-field evolution. The approach is illustrated for a system of particles governed by a pairing Hamiltonian. 
\end{abstract}


\maketitle

Under certain conditions, it is possible to provide an approximate description for quantal evolution of a system in terms of  an ensemble of classical trajectories with proper choice of initial conditions. This aspect appears naturally in the path integral formulation of quantum dynamics and has been recognized in refs. \cite{Her84,Kay94}. In recent years, this idea has been pushed forward to improved the mean-field description of many-body interacting systems \cite{Ayi08,Lac04}. Mean-field description cannot describe essential quantal effects associated to collective motion and severely underestimates fluctuations of collective observables. By considering an ensemble of mean-field trajectories with a specific choice of the fluctuations (quantal zero-point and thermal) in the initial state,  it is possible to overcome some of these shortcomings. Several applications, especially in the nuclear physics context \cite{Ayi09,Was09,Ayi10,Yil11}, have shown that this approach can improve the mean-field description by including important dissipative aspects in transport properties of heavy-ion collisions.  More recently, we illustrated that such an approach, in addition to fluctuations,  can accurately tackle the problem of symmetry breaking close to bifurcation point in collective energy landscape \cite{Lac12}.   Up to now, the stochastic mean-field (SMF) approach with initial fluctuations has been developed starting from a Time-Dependent Hartree-Fock (TDHF) version of the mean-field. Nowadays, there are  increasing interests in the treatment of pairing to describe evolution of strongly interacting Fermi liquids \cite{Ton06,Ave08,Eba10,Ste11,Sca13} employing the Time-Dependent Hartree-Fock Bogoliubov (TDHFB)  approach or its simplified BCS limit. 
While the TDHFB theory provides an important improvement beyond TDHF, it still suffers from the above quoted limitations: i.e. underestimation of quantum collective fluctuations and impossibility to spontaneously break symmetries.
Due to the successful description of the SMF approach in geometric symmetry breaking, it is rather tempting to introduce quantum fluctuations through initial sampling to treat pairing where the $U(1)$ symmetry breaking plays an important role.

In this work, we present an extension of the SMF approach by incorporating pairing into the description. 
In the standard TDHFB theory, evolution of the generalized density matrix ${\cal R}(t)$ is given by 
\begin{eqnarray}
i \hbar \frac{d }{dt}{\cal R}(t)
&=& \left[{\cal H}({\cal R}), {\cal R}(t)  \right]. \label{eq:tdhfb}
\end{eqnarray} 
${\cal R}(t)$ contains both normal $\rho_{ij} = \langle a^\dagger_j a_i \rangle$ and anomalous 
$\kappa_{ij} = \langle a_j a_i \rangle$ density matrix components. Here, $(a^\dagger_i , a_i)$ are creation and annihilation operators of a given single-particle basis.  The quantity ${\cal H}({\cal R})$ is the generalized mean-field Hamiltonian  containing both the mean-field $h[\rho]$ and the pairing field $\Delta[\kappa]$ \cite{Bla86}. 
Eq. (\ref{eq:tdhfb}) is generally solved either starting from a 
density associated to a quasi-particle vacuum or a statistical ensemble of quasi-particles. It is then convenient to introduce associated quasi-particle creation operators, defined through 
$\gamma^\dagger_\alpha =  \sum_{i} U^*_{i \alpha} a_i + V^*_{i\alpha} a^\dagger_i$\footnote{In the following, Greek index will be reserved to quasi-particles while latin index will refers to particles.}.
In quasi-particle representation, normal density matrix is diagonal and expectation value of anomalous density matrix vanishes, 
\begin{eqnarray}
\rho_{\alpha \beta} = \langle \gamma_\beta \gamma_\alpha ^\dagger \rangle = \delta_{\alpha\beta} f_\alpha;  ~~~~
\kappa_{\alpha \beta} = \langle \gamma_\beta \gamma_\alpha \rangle = 0.
\end{eqnarray}
Note that these rho and kappa are not the usual ones, but defined in terms of the quasi-particle operators.
Here, $f_\alpha$ denotes the occupation numbers of quasi-particle states \cite{Bla86,Goo81} and the expectation values are taken with respect to the quasi-particle vacuum. For a quasi-particle statistical ensemble $f_\alpha = (1+\exp[-\beta E_\alpha])^{-1}$ where $E_\alpha>0$ are the 
quasi-particle energies. As a result,
${\cal R}(t)$ takes a diagonal form 
\begin{eqnarray}
{\cal R}(t) = \sum_\alpha \left| W_\alpha(t) \right>f_\alpha \left<  W_\alpha(t) \right|  +
\left| V_\alpha(t) \right> (1-f_\alpha) \left<  V_\alpha(t) \right|
 \label{eq:densdiag}
\end{eqnarray} 
where  vectors $| V_\alpha(t) \rangle$ and $| W_\alpha(t) \rangle$ are the eigenvectors  of ${\cal R}(t)$ whose expressions can be found in 
\cite{Bla86}. In the following, for notation compactness, we will use $| S_\alpha(t) \rangle$ for both types of states. Then, in the TDHFB 
dynamics, we need to evolve state vectors according to $i \hbar \partial_t |S_\alpha \rangle =  {\cal H}({\cal R}) | S_\alpha \rangle$,
while keeping the quasi-particle occupation fixed. 
In the stochastic extension of TDHFB, quantal zero-point fluctuations (and thermal fluctuations) in the initial state are incorporated into the description in a similar manner to the SMF approach developed in \cite{Ayi08}. Initial fluctuations are simulated by generating  an ensemble of initial density matrices.  Each member of the initial density matrix, labelled by $(n)$ is written as,
\begin{eqnarray}
{\cal R}^{(n)}(t_0) &=& \sum_{\alpha \beta } | S_\alpha(t_0) \rangle {\cal R}^{(n)}_{\alpha \beta} \langle S_\beta(t_0) |. \label{eq:rini}
\end{eqnarray} 
where summations over $\alpha$ and $\beta$ run over a complete set of state vectors. The statistical properties of the elements of density matrix 
${\cal R}^{(n)}_{\alpha \beta}(t_0)$ is specified in terms of statistical properties of the normal $\rho^{(n)}_{\alpha\beta}(t_0)$ and 
anomalous $\kappa^{(n)}_{\alpha\beta}(t_0)$ density matrices. Elements of the normal and the anomalous density matrices are uncorrelated 
Gaussian random numbers with the mean values,   
\begin{eqnarray}
\overline{\rho^{(n)}_{\alpha \beta} }&=& \delta_{\alpha\beta} f_\alpha, ~~~\overline{\kappa^{(n)}_{\alpha\beta}} = 0, \label{eq:flucrk}
\end{eqnarray} 
and the second moments defined by,
\begin{eqnarray}
\overline{ \delta \rho^{(n)}_{\alpha \beta} ~  \delta \rho^{(n)*}_{\alpha' \beta'}}  =  
\frac{1}{2}  \delta_{\alpha \alpha'} \delta_{\beta \beta'} \left[ f_\alpha (1 - f_\beta)  + f_\beta (1-f_\alpha) \right], \label{eq:flucrho} \\
\overline{ \delta \kappa^{(n)}_{\alpha \beta} ~  \delta \kappa^{(n)*}_{\alpha' \beta'} }  = 
\frac{1}{2}  \delta_{\alpha \alpha'} \delta_{\beta \beta'} \left[ f_\alpha f_\beta + (1-f_\alpha) (1-f_\beta) \right]. \label{eq:fluckappa}
\end{eqnarray}
These relations have been deduced using the strategy discussed in appendix \ref{app:a}.

In these expressions, $\delta \rho^{(n)}$ and $\delta \kappa^{(n)}$ are the fluctuating parts of the density matrices,
\begin{eqnarray}
\rho^{(n)} = \overline{\rho^{(n)} }+ \delta \rho^{(n)} , ~~~ \kappa^{(n)} = \overline{\kappa^{(n)} }+ \delta \kappa^{(n)},
\end{eqnarray} 
and notation $\overline{X}$ indicates the average over the initial ensemble. 
We note that in the SMF approach without pairing only fluctuations in the particle-hole channels are included. In the extension including pairing, additional fluctuations originating from particle-particle and hole-hole channels appears from Eqs. 
(\ref{eq:flucrho}-\ref{eq:fluckappa}). This is an important new aspect which allows  to explore initial conditions with non-zero anomalous densities.   

In the novel stochastic approach, an event of the ensemble of generalized density matrices is expressed as, 
\begin{eqnarray}
{\cal R}^{(n)}(t) &=& \sum_{\alpha \beta } | S^{(n)}_\alpha(t) \rangle {\cal R}^{(n)}_{\alpha \beta} (t_0) \langle S^{(n)}_\beta(t) |,
\end{eqnarray}  
where states of a given event evolve with their own self-consistent generalized Hamiltonian according to the TDHFB 
equation of motion, $i \hbar \partial_t |S^{(n)}_\alpha \rangle =  {\cal H}({\cal R}^{(n)}) | S^{(n)}_\alpha \rangle$. 
In the TDHFB framework, we can  consider generalized operators that can create or annihilate two particles, 
\begin{eqnarray}
\hat Q &=& \sum_{ij} Q^{11}_{ij} a^\dagger_i a_j + \sum_{ij}  \left( Q^{20}_{ij} a_j a_i + Q^{20*}_{ji}  a^\dagger_i a^\dagger_j \right).
\end{eqnarray}  
The expectation value of such an operator in each event generated in the stochastic TDHFB approach reads
\begin{eqnarray}
Q^{(n)}(t) = {\rm Tr}(Q^{11} \rho^{(n)}(t)) + 2 \Re {\rm Tr} (Q^{20} \kappa^{(n)} (t)) \label{eq:evolobs}.
\end{eqnarray}  
Then, we determine the average evolution $\overline{Q}(t)$ by taking the average over the ensemble generated in the simulations. 
Since the stochastic approach incorporates correlations beyond the mean-field description, the ensemble average value $\overline{Q}(t)$,
in general, is different than the expectation value of the operator in the standard TDHFB approach.  Dispersion of the observable
$\hat Q$ is calculated using the classical formula,
\begin{eqnarray}
\sigma^2_Q(t) &=& \overline{(Q^{(n)}(t) - \overline{Q}(t))^2} \label{eq:clas}.
\end{eqnarray}   
We note that  fluctuations (\ref{eq:flucrho}-\ref{eq:fluckappa}) are 
chosen to insure that the initial ensemble average of the first and second 
moments are equal to those obtained from the quasi-particle density matrix 
at zero or finite temperature.

It is rather interesting to mention that Eqs. (\ref{eq:flucrho}) and (\ref{eq:fluckappa}) provide fluctuations 
in the quasi-particle basis which give specific aspects. For instance, at zero temperature, Eq. (\ref{eq:flucrho})
cancels out and only (\ref{eq:fluckappa}) gives non-zero fluctuations. Therefore, although it  might appear surprising, the T=0 limit is fully contained in the fluctuation of the anomalous density especially in the case of vanishing pairing. As an illustration, let us consider that the initial state is a Slater determinant.  In that case quasi-particle creation operators can be either particle creation operators (with 
$f_\alpha = 1-n_\alpha$) or holes annihilation operators (with $f_\alpha = n_\alpha$), where $n_\alpha=1,0$ are the single-particle occupation number. Then Eq.  (\ref{eq:fluckappa}) identifies with the fluctuation of the one-body density originally proposed in ref.  \cite{Ayi08}.
In a similar way, the case of a statistical ensemble at finite temperature with zero pairing can be deduced. Then, both Eqs. (\ref{eq:flucrho}) and (\ref{eq:fluckappa}) contribute to fluctuations.

In order to illustrate the powerfulness of the approach, we consider a many-body system governed by a pairing Hamiltonian. In the model, there are $K$ single-particle levels, labelled by $i$. Each level is associated
with the energy $\varepsilon_i$ and  has a degeneracy $2 \Omega_i$. It is assumed that each time a state $i$ is present, its time-reversed states $\bar i$ is also  present with same energy. We define a number operator for each energy level as 
$\hat N_i = \sum_p^{\Omega_i} (a^\dagger_{p,i}  a_{p,i} + a^\dagger_{\bar p,i}  a_{\bar p,i})$. 
In addition, pair creation/annihilation operators $
\hat S^+_i  = \sum_p^{\Omega_i} a^\dagger_{p,i} a^\dagger_{\bar p,i}$, $\hat S^-_i  =   \left( \hat S^+_i \right)^\dagger$ are introduced.
The pairing Hamiltonian of the system is given by \cite{Ric64}
\begin{eqnarray}
H & = &  \sum_{i=1}^{K}  \varepsilon_i \hat N_i + 
\sum_{i j}^K  G _{ij} \hat S^+_i   \hat S^-_j.    \label{eq:rich}
\end{eqnarray}  
From the set of operators, one can construct 
three quasi-spin components $\hat S^x_i =  ( \hat S^+_i + \hat S^-_i )/2$, $S^y_i = ( \hat S^+_i - \hat S^-_i )/2i$
and $\hat S^z_i = (\hat N_i - \Omega_i)/2$, which form a standard SU(2) algebra. 
For not too large particle and single-particle levels numbers, 
the eigenstates of $H$ can be determined by direct diagonalization \cite{Vol01} giving access 
to the exact static and dynamical properties.

Within TDHFB, evolutions of the expectation values $(S_i^x(t),S_i^y(t),S_i^z(t))$ 
of the quasi-spin components are determined by a set of coupled equations ($\hbar=1$),
\begin{eqnarray}
\frac{d}{dt} 
\left( 
\begin{array} {c}
   S^x_i(t)  \\
   S^y_i(t)  \\
   S^z_i(t)
\end{array} 
\right) =  \left(
\begin{array} {c c c}
0 &  -2  \tilde \varepsilon_i (t)  &  + 2 \Delta^y_i  \\
2 \tilde \varepsilon_i (t) & 0 &  -2 \Delta^x_i  \\
- 2 \Delta^y_i  & 2 \Delta^x_i & 0
\end{array}
\right) 
\left( 
\begin{array} {c}
   S^x_i(t) \\
   S^y_i(t) \\
   S^z_i(t)
\end{array} 
\right) \label{eq:tdhfbspin}
\end{eqnarray}
where 
\begin{eqnarray}
\tilde \varepsilon_i (t) =  \varepsilon_i + G_{ii} + 2G_{ii} \frac{\langle S^z_i \rangle}{\Omega_i} 
\end{eqnarray}
is nothing but the self-consistent mean-field while the pairing field $\Delta^{x (y)}_{i}(t) = \sum_k G_{ik} S^{x (y)}_i(t)$. These equations of
motion are quite general. They appear for instance
in Fermi gas dynamics simulations on a lattice with a contact interaction \cite{Ton06} or in the 
nuclear physics context \cite{Eba10,Sca13}. 

In order to illustrate some limitations of TDHFB, let us  
suppose that the initial state of the system is described by a Slater determinant of single-particle 
wave functions. Since  the Hamiltonian (\ref{eq:rich}) contains two-body interactions, this
state is not an eigenstate of the pairing Hamiltonian and it is expected to evolve in time. However, 
it turns out that this state is a stationary solution of the TDHFB equation. Indeed, since the state 
is an eigenstate of the particle number, the expectation values of quasi-spins vanish $S^x_i(t_0)= S^y_i(t_0) = 0$.
Hence, we deduce that all quasi-spin components remain constant in time.  
Therefore, the TDHFB approach is unable to describe correlations that built up in time and 
leads to the departure from the independent particle approach. This failure stems from the fact that a symmetry 
that exists initially, here the $U(1)$ symmetry, cannot be spontaneously broken in the mean-field approach.  
In the stochastic TDHFB approach, the situation is different. Time evolution of the expectation values of 
quasi-spin components are still determined by the same set of equations (\ref{eq:tdhfbspin}). However, the 
initial condition is not determined in a deterministic manner, but is specified in terms of distributions of 
$(S^{x(n)}_i, S^{y(n)}_i, S^{z(n)}_i)_{i=1,K}$. Following from the basic postulate
of the approach, (\ref{eq:flucrho}) and (\ref {eq:fluckappa}), the initial distributions of quasi-spin components
are uncorrelated random Gaussian numbers. Their mean values are determined by 
$\overline{S^{x(n)}_i} = \overline{S^{y(n)}_i} = 0$ while $\overline{S^{z(n)}_i} = +1/2$ or $-1/2$ 
for occupied and unoccupied levels respectively. Identifying the quasi-particle vacuum with a Slater determinant, 
variances of quasi-spin distributions are given by,
\begin{eqnarray}
\overline{ S^{z(n)}_i S^{z(n)}_j } &=& \overline{ S^{z(n)}_i S^{x(n)}_j } = \overline{ S^{z(n)}_i S^{y(n)}_j }
= \overline{ S^{x(n)}_i S^{y(n)}_j } = 0   \nonumber
\end{eqnarray}  
and 
\begin{eqnarray}
\overline{ S^{x(n)}_i S^{x(n)}_j } &=& \overline{ S^{y(n)}_i S^{y(n)}_j }  = \frac{1}{4} \delta_{ij} \Omega_i. \nonumber
\end{eqnarray}
Consequently, for each level $i$, the $z$ component of quasi-spin is initially a non-fluctuating quantity, while the
$x$ and $y$ components ($S^{x(n)}_i,S^{y(n)}_i$) are specified by uncorrelated  real Gaussian distributions.  
Solution is not a deterministic final state, but consists of a distribution of final states.  
Note that the mean number of particle is conserved event by event while the total energy is conserved in average.
\begin{figure}[htbp] 
\includegraphics[width=8.5cm]{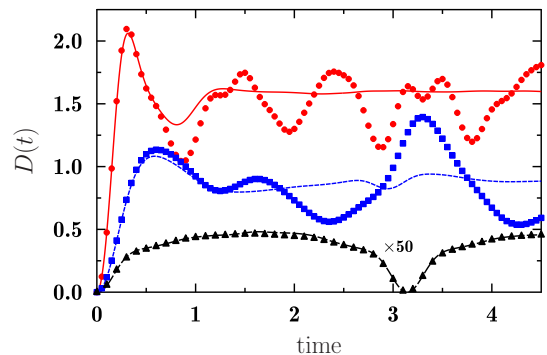} 
\caption{ (color online) Exact evolution of $D(t)$ for $G/\Delta \varepsilon = -0.05$ (black triangles), $-0.5$ (blue squares) and 
$-1$ (red circles). The results obtained by averaging over TDHFB trajectories are shown respectively by black long dashed line, 
blue short dashed line and red solid line. Note that curves corresponding to $G/\Delta \varepsilon = -0.05$ have been multiplied by 50. The time is presented in $(\Delta \varepsilon)^{-1}$ unit.} 
\label{fig:dt} 
\end{figure}

\begin{figure}[htbp] 
\vspace{0.8cm}
\includegraphics[width=8.5cm]{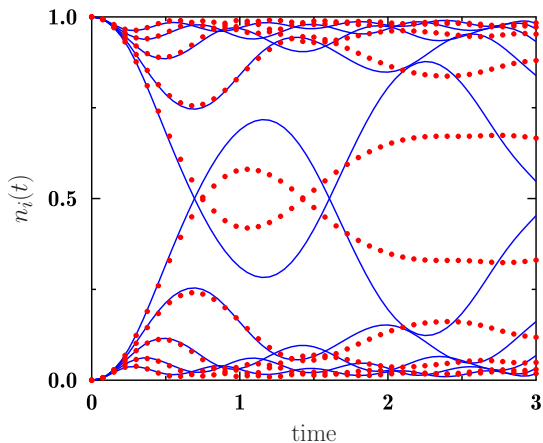} 
\caption{ (color online) Evolution of the occupation numbers as a function of time for the intermediate coupling strength
$G/\Delta \varepsilon = -0.5$. The exact occupation numbers are shown by solid lines while the SMF result is shown by filled circles.} 
\label{fig:nit} 
\end{figure}

In the illustrations below, we consider $K=10$ doubly-degenerated levels  ($\Omega_i = 1$) 
with constant level spacing $\Delta \varepsilon$ between adjacent levels, and constant pairing interaction $G_{ik} = G$.  
A system of $N=10$ particles occupy initially the five lowest energy levels. The exact evolution is obtained by decomposing 
the initial state in terms of eigenstates of the pairing Hamiltonian. The results are compared with stochastic simulations by
generating $N_{\rm evt} = 2\cdot 10^5$ number of events. We obtain solutions of Eq. (\ref{eq:tdhfbspin}) by employing a Runge-Kutta 2 
algorithm with a numerical time step $\Delta t = 0.005 / \Delta \varepsilon$. In Fig. \ref{fig:dt} the quantity $D(t) = {\rm Tr(\rho(t) - \rho^2(t))}$, 
which illustrates the time-scale associated to the departure from the independent particle picture, is shown 
as a function of time. In the canonical basis, this quantity can be expressed as: 
\begin{eqnarray}
D(t) &=&  \sum_i \Omega_i n_i(t) (1-n_i(t)) .
\end{eqnarray}   
Here $n_i(t)$ are the single-particle 
occupation numbers. From this expression, we see that $D(t)$ is zero if all occupation are equal to 0 or 1, i.e. when the state identifies
with the Slater determinant. 
In this figure, symbols  show the exact results and solid lines  denote the simulations of stochastic TDHFB
approach. In the stochastic TDHFB, the average occupation numbers are determined by the following ensemble averages 
$n_{i}(t) = 1/2 + \overline{ S^{z(n)}_i} /\Omega_i$. As an illustration, we also give in Fig. \ref{fig:nit} the occupation numbers 
evolution for $G/\Delta \varepsilon =-0.5$. From this figure, one can see that the agreement is very good over short time and then deviation are observed. The biggest deviation is seen for levels lying close to the initial Fermi energy.

The stochastic approach with pairing is able to incorporate non-trivial dissipative effects due to the coupling 
between complex internal degrees of freedom. Incorporating non-trivial initial fluctuations in particle-particle (pp) and 
hole-hole (hh) channels is able in all cases to describe short time dynamics very well. When the interaction strength increases, 
evolution of $D(t)$ becomes more complex and tends to oscillate around an average value. The stochastic approach is accurate 
not only for short time behavior, but also gives a quite reasonable description of the asymptotic behavior
although the evolutions appear more damped than the exact solutions. 
In the weak coupling regime $G/\Delta \varepsilon = -0.05$, the exact dynamics exhibits a periodic evolution with a rather short 
Poincar\'e recurrence time that is perfectly reproduced by the average TDHFB evolution. We note that, with the present parameter 
value for the Hamiltonian, the BCS threshold is at  $G_{\rm cr} / \Delta \varepsilon \simeq -0.3$, i.e. above this coupling strength, 
a $U(1)$ symmetry breaking  mean-field solutions exists. It is important to recall that the standard TDHFB gives $D(t)=0$. 
Similarly, the previously proposed  version of SMF approach with fluctuations on the normal density only and that was not exploring the possibility to break the particle number 
symmetry also leads to $D(t)=0$.  

\begin{figure}[htbp] 
\includegraphics[width=8.5cm.]{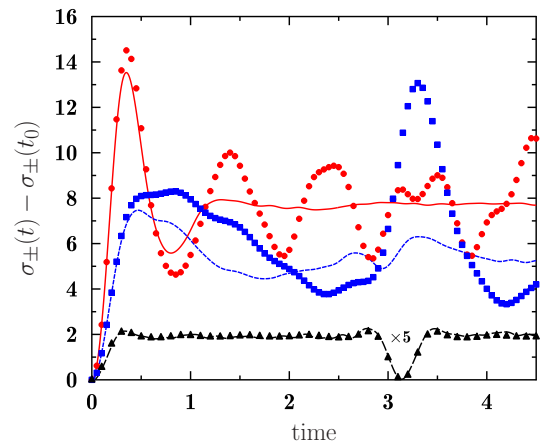} 
\caption{ (color online) Comparison of the exact and average evolution of $\sigma_{\pm}(t)$ for different coupling strengths. 
Conventions are the same as in Fig. \ref{fig:dt}.  Note that curves corresponding to $G/\Delta \varepsilon = -0.05$ have been 
multiplied by a factor 5.} 
\label{fig:sigma} 
\end{figure}

As another illustration, the two-body quantity $\sigma_{\pm}(t) = \sum_{ij} \langle \hat S^{+}_i \hat S^{-}_j  \rangle (t) $ is considered. 
Fig. \ref{fig:sigma} shows this quantity as a function of time for the exact and the stochastic TDHFB approach for different values of 
coupling strength $G/\Delta \varepsilon$. Symbols indicate  the exact evolutions and the solid lines are  the results of the stochastic
TDHFB simulations where fluctuations are computed from 
\begin{eqnarray}
\sigma_{\pm}  (t)&=& \sum_{ij} 
\left(\overline{S^{x(n)}_i(t) S^{x(n)}_j(t)} + \overline{S^{y(n)}_i(t)S^{y(n)}_j(t) } \right)?
\end{eqnarray}
Here also, we observe the stochastic approach reproduces perfectly the exact evolution of quantal fluctuations during relatively 
short time scale for all coupling strengths, and provides a reasonable description of the  gross properties (i.e. time average behavior) 
of the fluctuations over a long time scale for all coupling strengths.

\begin{figure}[htbp] 
\includegraphics[width=8.5cm.]{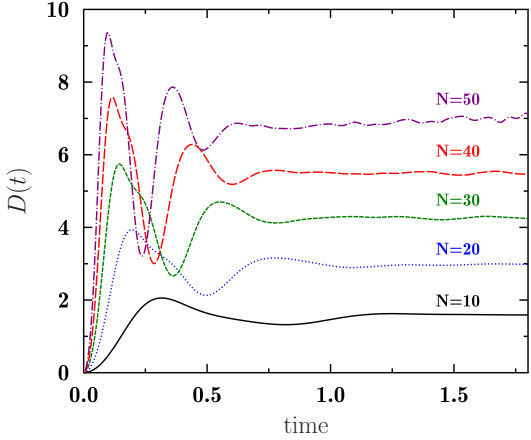} 
\caption{ (color online) Evolution of $D(t)$ for various number of particles: $N=10$ (solid line), $N=20$ (dotted line),  $N=30$ (short dashed line), $N=40$ (long dashed line) and $N=50$ (dot-dashed line). In all cases, $G/\Delta \varepsilon =-1$ and the number of levels $K=N$. Note that, the level spacing has been reduced as $N$ increases to insure that the maximal energy of the single-particle levels remains constant.} 
\label{fig:dtn} 
\end{figure}

Up to now, we have shown results for cases where the number of particles, i.e. number of degrees of freedom (DoF), is low enough
to allow for exact calculation. This was particularly useful to benchmark the present approach and shows that correlation beyond mean-field are properly included especially for short time dynamics. 
The real usefulness of the method will become obvious if the number of  DoF becomes large. In the present model, when $N \ge 16$, an exact calculation becomes prohibitive while the SMF approach is possible even for large 
number of particles. In Fig. \ref{fig:dtn}, examples of results obtained with SMF for various number of particles are shown. 
These calculations take few minutes on a standard laptop.    

In the present work, we propose a stochastic extension of the mean-field dynamics including pairing correlations. In this approach, quantal zero-point (and possibly thermal) fluctuations in the initial state are incorporated by simulating an ensemble of events by constructing a suitable distribution of the initial states. Observable quantities are determined by averaging  over the generated ensemble. The previous illustrations of the approach without pairing \cite{Ayi08,Lac12} and  the present application including the pairing correlations show that the stochastic approach reproduces a reasonable description of the gross properties exact quantal dynamics. The applications 
of SMF with or without pairing give a strong support for the fact the stochastic extension of the mean-field without and with pairing provide a  powerful tool for an approximate description of the correlations and quantal fluctuations of collective motion in normal and superfluid mesoscopic systems, beyond the mean-field description. The TDHFB approach has recently became a widely employed tool to describe Fermi liquids appearing in nuclear, condensed matter or atomic physics.  With rapid progress of computational powers, it is not anymore unreasonable to simulate thousands of independent TDHFB events, and consequently the stochastic approach proposed in this work  may provided as the next 
step to further progress in the description of transport properties of such physical systems.

\appendix

\section{Some remarks on Eq. (\ref{eq:flucrho}) and (\ref{eq:fluckappa})}
\label{app:a}

The initial fluctuations provided by Eqs. (\ref{eq:flucrho}) and (\ref{eq:fluckappa}) plays a crucial role in the 
present stochastic mean-field with pairing. The main idea behind the SMF approach is that the initial quantum fluctuations 
can be replaced by classical fluctuations in the collective space. Let us consider a one-body observable 
$\hat Q = \sum_{ij} \left< i \left| Q \right| j \right> a^\dagger_i a_j$. Its expectation value on a statistical quasi-particle ensemble 
are given by:
\begin{eqnarray}
\langle \hat Q \rangle &=&  \sum_{ij\alpha } \left< i \left| Q \right| j \right> \left\{ V_{i\alpha} f_\alpha V^*_{j\alpha }  + U^*_{i \alpha}  (1-f_\alpha) U_{j \alpha}  \right\} \label{eq:mean}
\end{eqnarray}
while its quantal fluctuation reads:
\begin{eqnarray}
\sigma^2_Q &=& \langle \hat Q \hat Q \rangle - \langle \hat Q \rangle \langle \hat Q \rangle \nonumber \\
&=& \sum_{ijkl\alpha \beta } \langle i |  \hat Q | j \rangle \langle k |  \hat Q | l \rangle V_{i\alpha} V^*_{j\beta }  V_{k\beta} V^*_{l\alpha }  f_\alpha (1 - f_\beta)  \nonumber \\
&&+ \sum_{ijkl\alpha \beta } \langle i |  \hat Q | j \rangle \langle k |  \hat Q | l \rangle U^*_{i \alpha}  U_{j \beta}  U^*_{k \beta}  U_{l \alpha}  
 f_\beta (1-f_\alpha) \nonumber \\
 &&- \sum_{ijkl\alpha \beta } \langle i |  \hat Q | j \rangle \langle k |  \hat Q | l \rangle U^*_{i \alpha}  U_{j \beta}  V_{k\alpha} V^*_{l\beta} f_\beta(1-f_\alpha) \nonumber \\
&&- \sum_{ijkl\alpha \beta } \langle i |  \hat Q | j \rangle \langle k |  \hat Q | l \rangle V_{i\alpha} V^*_{j\beta } U^*_{k \alpha}  U_{l \beta} f_\alpha (1-f_\beta) \nonumber \\
&&+\sum_{ijkl \alpha\beta}   \langle i |  \hat Q | j \rangle \langle k |  \hat Q | l \rangle U^*_{i \alpha} V^*_{j\beta}   V_{k\beta}  U_{l \alpha}  \nonumber \\
&&\hspace*{1.cm} \times \left[ f_\alpha f_\beta + (1-f_\alpha) (1-f_\beta) \right] \nonumber \\
&&- \sum_{ijkl  \alpha\beta}   \langle i |  \hat Q | j \rangle \langle k |  \hat Q | l \rangle U^*_{i \alpha} V^*_{j\beta}   V_{k\alpha}  U_{l \beta}
\nonumber \\
&&\hspace*{1.cm} 
\times \left[ f_\alpha f_\beta + (1-f_\alpha) (1-f_\beta) \right] \label{eq:fluc}
\end{eqnarray}
    
We now consider a set of initial value $Q^{(n)}$ written as:
\begin{eqnarray}
Q^{(n)} &=& \sum_{ij\alpha\beta } \langle i |Q| j \rangle
 \left[  V_{i\alpha} V^*_{j\beta }  \langle  \gamma_\alpha \gamma^\dagger_\beta \rangle^{(n)} 
+ U^*_{i \alpha}  U_{j \beta}  \langle\gamma^\dagger_\alpha \gamma_\beta  \rangle^{(n)} \right. \nonumber \\
&& \left. +   U^*_{i \alpha} V^*_{j\beta}   \langle \gamma^\dagger_\alpha \gamma^\dagger_\beta  \rangle^{(n)}
+    V_{i\alpha}  U_{j \beta}  \langle \gamma_\alpha \gamma_\beta \rangle^{(n)} \right] \label{eq:qn}
\end{eqnarray}  
where the quantity $\langle  \gamma_\alpha \gamma^\dagger_\beta \rangle^{(n)} $, $\langle\gamma^\dagger_\alpha \gamma_\beta  \rangle^{(n)} $, $\langle \gamma^\dagger_\alpha \gamma^\dagger_\beta  \rangle^{(n)}$ and $ \langle \gamma_\alpha \gamma_\beta \nonumber  \rangle^{(n)}$ are Gaussian random variables whose properties are chosen to insure that the mean-value and fluctuation of the classical 
variable $Q^{(n)}$ equal the quantum expectation value given respectively by $(\ref{eq:mean})$ and $(\ref{eq:fluc})$. We see that the conditions:
\begin{eqnarray}
 \overline{ \langle \gamma_\alpha \gamma^\dagger_\beta \rangle  ^{(n)} }
&=& 1 - \overline{ \langle \gamma^\dagger_\alpha \gamma_\beta \rangle^{(n)} } = \delta_{\alpha \beta } f_\alpha , \nonumber \\
 \overline{ \langle \gamma_\alpha \gamma_\beta \rangle ^{(n)} } &=&
\overline{ \langle \gamma^\dagger_\alpha \gamma^\dagger_\beta \rangle^{(n)} } = 0 \nonumber
\end{eqnarray}
automatically insure $\overline{Q} =  \langle \hat Q \rangle$. These conditions are nothing but Eqs. (\ref{eq:flucrk}).

We now introduce the fluctuation around the mean-values, i.e. for a variable $X$, the quantity $\delta X = X^{(n)} - \overline{X^{(n)}}$. The initial fluctuations of different quantities entering in  Eq. (\ref{eq:qn}) have been obtained using the following 
strategy. Having in mind the original SMF approach without pairing, and assuming that quasi-particles plays now the role 
of particle in standard SMF, leads us to postulate Eq. (\ref{eq:flucrho}). In addition, it has been assumed that fluctuations where 
quasi-particle creation (resp. annihilation) operators appear one or three times exactly cancel out, i.e.:
\begin{eqnarray}
\overline{ \delta \langle \gamma^\dagger_\alpha \gamma^\dagger_\beta \rangle^{(n)} ~ \delta \langle \gamma^\dagger_\delta \gamma_\lambda \rangle^{(n)} }  & = & 0 \nonumber \\
\overline{ \delta \langle \gamma_\alpha \gamma_\beta \rangle^{(n)} ~ \delta \langle \gamma^\dagger_\delta \gamma_\lambda \rangle^{(n)} }  & = &  0 \nonumber 
\end{eqnarray} 
With these condition, we obtain that the classical fluctuations of $Q^{(n)}$ reads:
\begin{eqnarray}
\overline{\delta Q^{(n)} \delta Q^{(n)}} 
&=& \sum_{ijkl\alpha \beta } \langle i |  \hat Q | j \rangle \langle k |  \hat Q | l \rangle V_{i\alpha} V^*_{j\beta }  V_{k\beta} V^*_{l\alpha }  f_\alpha (1 - f_\beta)  \nonumber \\
&+& \sum_{ijkl\alpha \beta } \langle i |  \hat Q | j \rangle \langle k |  \hat Q | l \rangle U^*_{i \alpha}  U_{j \beta}  U^*_{k \beta}  U_{l \alpha}  
 f_\beta (1-f_\alpha) \nonumber \\
 &-& \sum_{ijkl\alpha \beta } \langle i |  \hat Q | j \rangle \langle k |  \hat Q | l \rangle U^*_{i \alpha}  U_{j \beta}  V_{k\alpha} V^*_{l\beta} f_\beta(1-f_\alpha) \nonumber \\
 &-& \sum_{ijkl\alpha \beta } \langle i |  \hat Q | j \rangle \langle k |  \hat Q | l \rangle V_{i\alpha} V^*_{j\beta } U^*_{k \alpha}  U_{l \beta} f_\alpha (1-f_\beta) \nonumber \\
&+& \sum_{ijkl \alpha\beta \gamma\delta } \langle i |  \hat Q | j \rangle \langle k |  \hat Q | l \rangle U^*_{i \alpha} V^*_{j\beta}  V_{k\lambda}  U_{l \delta}  \nonumber \\
&&\hspace*{1.cm} 
\times
\overline{ \delta \langle \gamma^\dagger_\alpha \gamma^\dagger_\beta  \rangle^{(n)}
\delta \langle  \gamma_\lambda \gamma_\delta  \rangle^{(n)} } \nonumber \\
&+& \sum_{ijkl \alpha\beta \gamma\delta } \langle i |  \hat Q | j \rangle \langle k |  \hat Q | l \rangle
V_{i\alpha}  U_{j \beta} U^*_{k \lambda} V^*_{l\delta} 
\nonumber \\
&&\hspace*{1.cm} 
\times 
\overline{\delta \langle \gamma_\alpha \gamma_\beta  \rangle^{(n)} \delta \langle \gamma^\dagger_\lambda \gamma^\dagger_\delta \rangle^{(n)}} \nonumber
\end{eqnarray} 
Comparing this expression with the equation (\ref{eq:fluc}) leads to the conclusion that a convenient choice of the fluctuation is:
\begin{eqnarray}
\overline{\delta \langle \gamma_\alpha \gamma_\beta  \rangle^{(n)} \delta \langle  \gamma^\dagger_\lambda
 \gamma^\dagger_\delta \rangle^{(n)}} & = & 
\overline{\delta \langle \gamma^\dagger_\alpha \gamma^\dagger_\beta  \rangle^{(n)} \delta \langle  
\gamma_\lambda \gamma_\delta \rangle^{(n)}} \nonumber \\
&=&
\frac{1}{2}
\left( \delta_{\alpha \delta }  \delta_{\beta \lambda} - \delta_{\alpha \lambda }  \delta_{\beta \delta} \right)  \nonumber \\
& & \times \left[ f_\alpha f_\beta + (1-f_\alpha) (1-f_\beta) \right] \nonumber
\end{eqnarray}   
That is consistent with Eq. (\ref{eq:fluckappa}).

\section*{Acknowledgment}  
S. A. gratefully acknowledge GANIL and IPN Orsay for hospitality extended to him during his visits. This work is supported in part by the US DOE Grant No. DE-FG05-89ER40530. We would also like to thank G. Scamps for discussions.


\begin{thebibliography} {99}
\bibitem{Her84} M. F. Herman and E. Kluk, Chem. Phys. {\bf 91}, 27 (1984).
\bibitem{Kay94} K. G. Kay, J. Chem. Phys. {\bf 100}, 4432 (1994); {\bf 101}, 2250 (1994).
\bibitem{Ayi08} S. Ayik, Phys. Lett. {\bf B658}, 174 (2008). 
\bibitem{Lac04} D. Lacroix, S. Ayik, and Ph. Chomaz, Prog. Part. Nucl. Phys.
{\bf 52}, 497 (2004).

\bibitem{Ayi09} S. Ayik, K. Washiyama, and D. Lacroix, Phys. Rev. {\bf C79}, 054606
(2009). 
\bibitem{Was09} K. Washiyama, S. Ayik, and D. Lacroix, Phys. Rev. {\bf C80},
031602(R) (2009). 
\bibitem{Ayi10} S. Ayik, B. Yilmaz and D. Lacroix, Phys. Rev. {\bf C81}, 034605 (2010).
\bibitem{Yil11} B. Yilmaz, S. Ayik, D. Lacroix, and K. Washiyama, Phys. Rev.
{\bf C83}, 064615 (2011).

\bibitem{Lac12} D. Lacroix, S. Ayik and B. Yilmaz, Phys. Rev. {\bf  C 85}, 041602(R) (2012). 


\bibitem{Ton06} G. Tonini, F. Werner and Y. Castin, Eur. Phys. J. D {\bf 39}, 283 (2006).

\bibitem{Ave08}  B. Avez, C. Simenel, and Ph. Chomaz, Phys. Rev. {\bf C 78}, 044318
(2008).

\bibitem{Ste11} I. Stetcu, A. Bulgac, P. Magierski, and K. J. Roche
Phys. Rev. {\bf C 84}, 051309, (2011).

\bibitem{Eba10} S. Ebata, T. Nakatsukasa, T. Inakura, K. Yoshida, Y. Hashimoto,
and K. Yabana, Phys. Rev. {\bf C 82}, 034306 (2010).

\bibitem{Sca13} G. Scamps and D. Lacroix, Phys. Rev. {\bf C87}, 014605 (2013).  


\bibitem{Bla86} J. P. Blaizot and G. Ripka, {\it Quantum Theory of Finite Systems}, (MIT Press, Cambridge,  
Massachusetts, 1986).  


\bibitem{Goo81} A.L. Goodman, Nucl. Phys. {\bf A352}, 30 (1981), A.L. Goodman, Phys. Rev. {\bf C29}, 1887 (1984).
 
\bibitem{Ric64} R. W. Richardson and N. Sherman, Nucl. Phys. {\bf 52}, 221 (1964);
R.W. Richardson, Phys. Rev. {\bf 141}, 949 (1966); J.Math. Phys. {\bf 9},
1327 (1968).

\bibitem{Vol01} A. Volya, B. A. Brown, and Z. Zelevinsky, Phys. Lett. {\bf B 509},
37 (2001); T. Sumaryada and Alexander Volya, Phys. Rev. {\bf C 76},
024319 (2007).


\end{thebibliography}
\end{document}